# Ultrasound Characterization of Oral Soft Tissues *in vivo* Using the Burr Speckle Model

Daria Poul
Dept. of Radiology
University of Michigan
Ann Arbor, MI, USA
ssheykho@umich.edu

Ankita Samal
Dept. of Radiology
University of Michigan
Iowa City, IA, USA
ankita-samal@uiowa.edu

Amanda Rodriguez Betancourt
Dept. of Periodontics
University of Illinois Chicago
Chicago, IL, USA
arodr368@uic.edu

Carole Quesada
Dept. of Radiology
University of Michigan
Ann Arbor, MI, USA
cquesada@umich.edu

Hsun-Liang Chan
Dept. of Periodontology
Ohio State University
Columbus, OH, USA
chan.1069@osu.edu

Oliver D. Kripfgans
Dept. of Radiology
University of Michigan
Ann Arbor, MI, USA
greentom@umich.edu

*Abstract*— **Periodontal (gum) diseases, reportedly affect 4 out of 10 adults 30 years of age or older in the USA. The standard of care for clinical assessments of these diseases is bleeding on probing, which is invasive, subjective and semi-qualitative. Thus, research on proposing alternative noninvasive modalities for clinical assessments of periodontal tissues is crucial. Quantitative Ultrasound (QUS) has shown promises in noninvasive assessments of various diseases in soft biological tissues; however, it has not been employed in periodontology. Here as the first step, we focused on QUS-based characterization of two very adjacent oral soft tissues of alveolar mucosa and attached gingiva in an *in vivo* animal study. We investigated first order ultrasonic speckle statistics using the two-parameter Burr model (power-law *b* and scale factor *l*). Our QUS analysis was compared with the Masson's Trichrome histology images of the two oral tissue types quantitatively using the RGB color thresholding. QUS study included 10 swine and US scanning was performed at the first and second molars of all four oral quadrants in each swine, resulting in 80 scans. US scan data was acquired at the transit/receive frequency of 24 MHz using a toothbrush-sized transducer. Parametric imaging of Burr parameters was created using a sliding kernel method with linear interpolations. The kernel size and overlap ratio was 10 wavelengths and 70%, respectively. No statistically significant difference was reported for estimated parameters when interpolation was performed (*p-value*>0.01). Results at both oral sites (molar 1 and molar 2) showed that the difference between the two tissue types using Burr parameters were statistically significant (*p-value*<0.0001). The average Burr *b* was reported to be higher in attached gingiva while the average Burr *l* was lower compared to mucosa. Visual comparison of Masson Trichrome histology images of these tissues showed denser color density in gingiva. The color thresholding of these images further confirmed that the percent of blue, which stains collagenous regions, was at least two times higher in gingiva than alveolar mucosa, based on threshold values. Comparing histology and QUS in characterizing the two tissue types, it was suggested that the elevated Burr *b* (related to potential scatterer densities) in gingiva could be aligned with findings from Masson's Trichrome histology. This study showed a promising potential of QUS for periodontal soft tissue characterization.**

## I. Introduction

Periodontal diseases reportedly affect 4 out of 10 adults (30 years of age or older) in the USA [1]. These diseases mainly result from inflammation and infection within different components of oral soft tissues such as alveolar mucosa, marginal (free) gingiva and attached gingiva. A schematic illustration of oral soft and hard tissues is shown in Fig. 1. The gingiva (marginal and attached) is a dense fibrous connective tissue with an intracellular structure for bearing loads (for example during mastication). The main role of gingiva is protecting the root and alveolar bone from being deformed or degraded. On the contrary, alveolar mucosa is mainly non-keratinized and has less exposure to abrasive forces. For clinical assessments of diseases in oral soft tissues, the standard of care is bleeding on probing (BOP). Here, a thin metallic probe is pushed into the pocket between teeth and marginal gingiva and parameters such as the depth of probing are recorded and ascribed to inflammation. However, BOP is invasive, subjective and semi-quantitative [2]. Thus, investigating noninvasive, objective and quantitative biomarkers for clinical assessments of periodontal tissues is crucial. Ultrasonic (US) imaging, as a noninvasive modality, has been employed in periodontology in the form of B-mode (brightness-mode) imaging for lesion detection, for measuring gingival thickness and also for delineating hard tissue surfaces (bone/crown) [3], [4]. Moreover, US-based elasticity estimations of oral soft tissues have also been studied in some studies [5]. Quantitative ultrasound (QUS) has shown promises in characterizations of various soft tissues [6], however, in periodontology QUS has only been employed in limited studies such as [7] where the employed method deviates from standard QUS analysis approaches. QUS analysis aims at obtaining fundamental properties of tissues from analyzing the interaction of ultrasound wave with tissue microstructures during US imaging [8]. For instance, the number and spatial positioning of tissue scatterers affect the acquired US data and the granular texture (a.k.a. speckle pattern) appeared on the B-mode image. It is caused by the interference of backscattered echoes from sub-resolution scatterers. First-order statistical modeling of US speckle could allow extracting quantitative features not visible on B-mode images [9]. In this study, we aim at characterizing two types of oral soft tissues (alveolar mucosa and gingiva) using QUS analysis based on the Burr speckle model in a swine model. We investigate the histology of these tissues, as well. Here, characterization of periodontal tissues at baseline contributes to an enhanced understanding of how difference in ultrastructure of oral tissue affects QUS. It serves as an initial step towards more advanced implementation of QUS analysis for periodontal inflammation diagnosis.

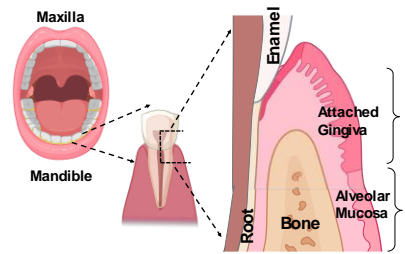

Fig. 1. A schematic illustration of oral soft and hard tissues.

## II. Theory

The first order US speckle statistics is basically the probability density function ($P(A)$) of the US echo envelope (A). It is assumed that changes in speckle model parameters could reflect and characterize changes in tissue pathology. Various speckle models have been employed for US speckle statistics to characterize soft tissues [9]. Here, we focus on a recently proposed framework leading to the Burr speckle model [10]–[12]. One of the main assumptions under this framework is presence of multi-scale power-law scatterers within tissues with respect to their characteristic sizes. According to (1), Burr distribution incorporates two parameters: a power-law key parameter $b$ (related to scatterer density) and a scale factor $l$ (related to echo intensity). Both parameters have shown potential in characterizing the soft tissues [13], [14].

$$P(A) = \frac{2A(b-1)}{l^2((A/l)^2 + 1)^b} \quad (1)$$

To estimate Burr parameters locally within an estimation kernel, we employ moment-based estimators in (2) and (3) [15]. Here, $E(A)$ is the first moment and $E[A^2]$ is the second moment (non-centered) of the echo amplitude.

$$E(A) = \frac{(b-1)l\sqrt{\pi}\Gamma(b-3/2)}{2\Gamma(b)} \quad (2)$$

$$\frac{E[A]^2}{E[A^2]} = \frac{(b-2)\pi(\Gamma(b-3/2))^2}{4(\Gamma(b-1))^2} \quad (3)$$

Funding Agency: National Institute of Dental and Craniofacial Research (1-R21- DE029005).

## III. Method

The swine study included intraoral US scanning of 10 pigs at their four first molars (M1) and four second molars (M2), i.e., left and right mandibular (MAND) and maxillary (MAX) oral sites, resulting in 80 scans. Ultrasonic IQ data were acquired at the transmit/receive frequency of 24 MHz by employing a clinical scanner (ZS3, Innovation Center, San Jose, CA, USA) equipped with a newly available small form factor linear array transducer (L30-8) [16]. The corresponding ROIs for QUS analysis of gingival and mucosal tissues were outlined on B-mode images to be devoid of the epithelium layer, bone, crown, or any heterogeneous region visually distinguishable in B-mode images. To estimate Burr parameters locally within a selected ROI, a sliding window method was employed with the kernel size of 10 wavelengths (0.064 mm) and the overlap ratio of 70% [17]. To create a smoother parametric image, linear interpolations between parameter estimations at adjacent estimations were performed. The potential effect of interpolation on parameter estimations was assessed using quantile-quantile (QQ) plots and the significance test (statistical pair-wise comparison using the nonparametric Mann-Whitney U test). For the histology analysis of gingiva and alveolar mucosa, Masson's Trichrome stain method was employed. In Masson's Trichrome stain, blue represents collagen fibers and other connective tissues. High-resolution (x20) microscopy images of histology slides were obtained by an optical microscope (E800, Nikon Instruments Inc., Melville, NY). Image processing was performed on histology images via a color thresholding method on the RGB (red, green, and blue) color space. In the RGB color model, each image pixel is composed of these color channels and color intensity values for each channel range from 0 to 255, where 0 indicates no contribution of that color and 255 means its maximum contribution. Our goal was to isolate blue in Masson's Trichrome images of gingiva and alveolar mucosa and compare them. Thresholding provided segmenting pixels with a blue pixel intensity higher than a selected threshold value while red and green pixel intensities were both below that threshold. We compared relative blue pixel percentages within ROIs in gingiva and mucosa.

## IV. Results and discussion

Fig. 2 demonstrates the effect of linear interpolation on parametric images for Burr $b$ compared with the case without interpolation for an ROI in a phantom. In Fig. 3 (a) and (b), QQ plots compare statistical distributions of estimations using interpolation vs. no interpolation for Burr $b$ and Burr $l$, respectively. Datasets from the two approaches closely lie along a 45-degree line, suggesting that they have similar statistical distributions. Also, *p-values* between estimations from interpolation and no interpolation were 0.29 and 0.28 for Burr $b$ and $l$, respectively, indicating that the linear interpolation did not cause any statistically significant difference in the estimated model parameters. For all 80 distinct oral scans, parametric images of Burr $b$ and $l$ were obtained for gingiva and alveolar mucosa. Fig. 4 shows a sample of US B-mode image of oral and hard soft tissues at the left mandibular oral site for molar 2 with the Burr parametric images. In the B-mode image, some important landmarks can be located such as the epithelium layer, appearing as a highly keratinized region covering gingiva, the bone (hyperechoic curved region underneath soft tissues) and the root (hyperechoic angled regions originated from the tooth at the left side). Local variations in estimated parameters are observed in oral soft tissues which suggest potential sensitivity of Burr parameters to local variations in ultrastructure of oral tissues. In Fig. 5, Burr parameters are compared in gingiva versus alveolar mucosa for molar 1 (top row) and molar 2 (bottom row) oral sites, each incorporating all respective scans. *P-values* are also reported for each pair. The difference between the two tissue types is statistically significant ($p\text{-}values<0.0001$) from the Burr parameter estimations. As observed consistently for the two tooth sites (M1 and M2) over the whole swine population, Burr $b$ (a measure of scatterer density) is elevated in gingiva compared to alveolar mucosa. On the other hand, Burr $l$ is reported to be significantly lower in gingiva in both cases. Thus, it is suggested that the Burr model could characterize the two periodontal soft tissues.

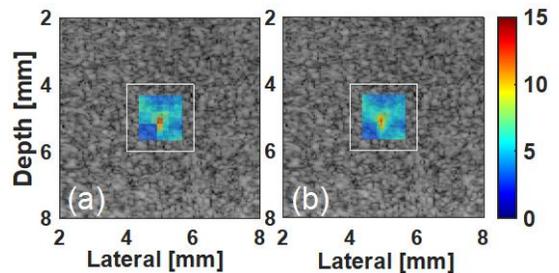

Fig. 2. Parametric imaging of Burr $b$ for a phantom scan. (a) without interpolations. (b) with the linear interpolation.

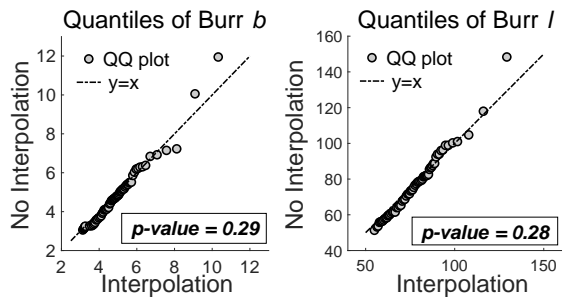

Fig 3. QQ plots comparing the effect of interpolation on parametric imaging in a phantom scan. (a) Burr $b$ and (b): Burr $l$.

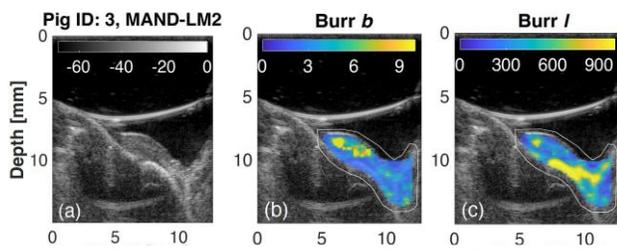

Fig. 4. (a) US image of periodontal tissues at left mandibular molar 2 site. (b) and (c) shows parametric images of Burr $b$ and Burr $l$, respectively.

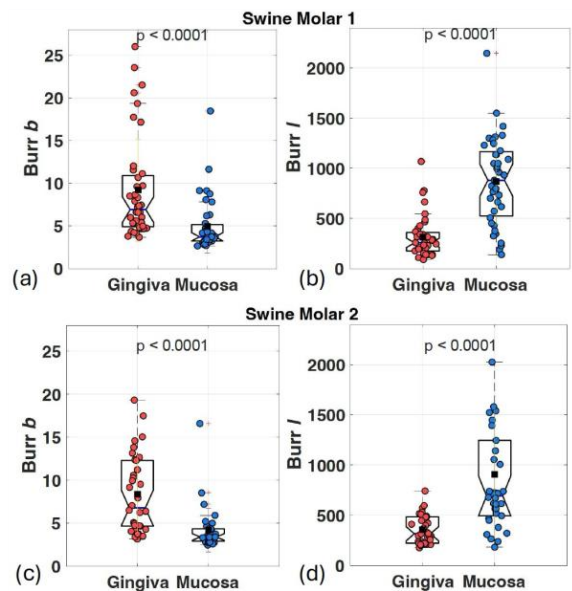

Fig. 5. Comparison of gingiva versus alveolar mucosa using Burr parameters ($b$ and $l$) for two oral sites of molar 1 ((a) and (b)) and molar 2 ((c) and (d)). *P-values* are also reported.

### A. Histology

To investigate potential difference in histology of these two periodontal tissues, a sample of histology image using Masson's Trichrome stain is compared in Fig. 6. The rectangular ROI shown as white box on (b) and (c) are selected for the histology assessment here. Visually, our high-resolution (x20 magnification) histology images of gingiva and alveolar mucosa show a difference between color density of stained tissues. Blue color appears denser in gingiva compared to mucosa. We have also compared histology images of the two tissues from a quantitative approach using RGB color thresholding, as explained in the Method section. In performing thresholding, the intensity threshold (cut-off) value needs to be selected. We tested threshold values ranging from 50 to 200 (increments of 10). For each case, we visually inspected segmented blue masks for the two tissue types and made a pair-wise analogy with their corresponding histology images to select reasonable threshold values for segmenting blue (collagenous) regions. To further elaborate our thresholding analysis, in Fig. 7 we have presented results for two different threshold values of 100 (top row), 140 (bottom row), comparing gingival and mucosal regions. The regional percentage of blue for each case is also reported on the upper side of the segmented blue masks. It is observed that for these segmentations, the fraction of blue in gingiva was at least twice greater than in alveolar mucosa. This histology findings in differentiation between these two tissue types could be aligned with QUS results for Burr $b$, which is associated with the potential scatterer density [10]. In QUS analysis, Burr $b$ was reported higher in gingiva compared to alveolar mucosa.

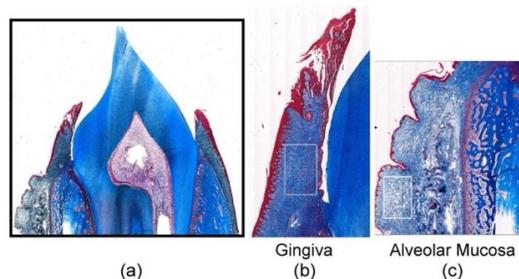

Fig. 6. (a) Masson's Trichrome histology image of swine oral tissues. (b) and (c): high-resolution microscopy imaging of gingiva and mucosa, respectively.

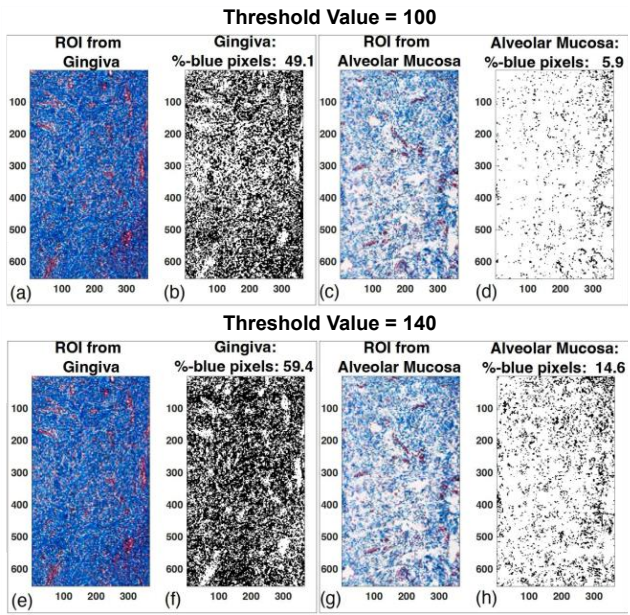

Fig. 7. Comparing gingiva (first column) vs. mucosa (third column) histology images using RGB thresholding. Threshold values: top row: 100, bottom row: 140. The second and fourth columns show blue channel masks for segmenting gingival and mucosal collagenous regions, respectively. % of blue obtained from RGB thresholding are also reported. ROIs from Fig. 6 are used.

## V. Conclusion

This study is among early investigations into applications of QUS and US speckle statistics for periodontal soft tissue characterization. We showed that Burr parameters could characterize gingiva versus alveolar mucosa in 80 swine scans. Findings from histology analysis comparing these two tissue types were suggested to be aligned with our QUS analysis. QUS holds potential in clinical assessment of periodontal soft tissues. Future studies should focus on periodontal inflammation characterization using QUS-based approaches.

## Acknowledgment

The authors acknowledge grant funding from the National Institute of Dental and Craniofacial Research (1-R21-DE-029005). The authors also thank Dr Mario Fabiilli for the loan of the microscopy imaging system.